\documentclass{aa}

\usepackage{graphicx}

\usepackage{epstopdf}
\usepackage{txfonts}
\usepackage[normalem]{ulem}
\usepackage{color}
\usepackage{xcolor, cancel}

\begin{document}

   \title{From Solar to Stellar Brightness Variations:}

   \subtitle{The Effect of Metallicity }

   \author{V.~Witzke\inst{1}\fnmsep\thanks{e-mail: witzke@mps.mpg.de}
          \and
          A.~I.~Shapiro\inst{1}
          \and
          S.~K.~Solanki\inst{1,2}
          \and          
          N.~A.~Krivova\inst{1}          
          \and
          W.~Schmutz\inst{3}
          }

   \institute{Max Planck Institute for Solar System Research, Justus-von-Liebig-Weg 3, 37077 G\"ottingen, Germany\\
         \and
         School of Space Research, Kyung Hee University, Yongin, Gyeonggi, 446-701, Republic of Korea\\
         \and
             Physikalisch-Meteorologishes Observatorium Davos, World Radiation Center, 7260 Davos Dorf, Switzerland\\
             }


 
  \abstract
 {Comparison studies of  Sun-like stars  with the Sun suggest an anomalously  low photometric variability of the Sun compared to Sun-like stars with similar magnetic activity.  Comprehensive understanding of stellar variability is needed, to find a physical reasoning for this observation.  }
   {We investigate the effect of metallicity and effective temperature on the photometric brightness change of Sun-like stars seen at different inclinations. The considered range of fundamental stellar parameters is sufficiently small so the stars, investigated here, still count as Sun-like or even as solar twins.}
   {To model the brightness change of  stars with solar magnetic activity,  we extend a well established  model of solar brightness variations, SATIRE (which stands for Spectral And Total Irradiance Reconstruction), which is based on solar spectra,  to stars with different fundamental parameters. For that we calculate stellar spectra for different metallicities and effective temperature using the radiative transfer code ATLAS9. }
   {We show that even a small change (e.g. within the observational error range) of metallicity or effective temperature significantly affects the photometric brightness change compared to the Sun. We find that for Sun-like stars, the amplitude of the brightness variations obtained for Str\"omgren {(\it{b} + \it{y})}/2    reaches a local minimum for fundamental stellar parameters close to the solar metallicity and effective temperature.  Moreover, our results show that the effect of inclination decreases for metallicity values greater than the solar metallicity. Overall, we find that an exact determination of fundamental stellar parameters is crucially important  for understanding stellar brightness changes.}
   {}


   \maketitle
%

\section{Introduction}
Stellar activity is caused by magnetic fields emerging from below the stellar surface and evolving due to the complex interaction of gas dynamics and magnetic flux. The resulting  concentrations of magnetic fields produce a variety of phenomena, including spots and faculae that lead to darkening and brightening on the stellar surface. Such emerging magnetic features  can be directly observed on the Sun, due to its exclusive location,  and thus have been  studied in great detail \citep[see, e.g.,][]{2006RPPh...69..563S}. In contrast, stellar activity can be  accessed only by indirect manifestations of the surface structures, i.e.  spectroscopic and brightness variations, as well as proxies of magnetic heating of the stellar atmosphere, chromospheric Ca~II and coronal X-ray emission. \\
Stellar long-term variability investigations were launched in the 50's by several observatories, such as the Lowell Observatory, to monitor variations of photometric brightness and chromospheric activity \citep[see][and references therein]{1978ApJ...226..379W, 1995ApJ...438..269B, 2013ASPC..472..203L}. These studies provide an observational survey of the relation between  photometric variations and chromospheric activity among lower main-sequence stars \citep{1998ApJS..118..239R,2007ApJS..171..260L, 2009AJ....138..312H, 2018ApJ...855...75R}, and led to several important findings. Firstly, the photometric variability is stronger for stars with higher magnetic activity. Secondly, it was found that among stars with a certain chromospheric activity level, a transition from faculae-dominated to spot-dominated stellar photometric variations occurs  \citep{2007ApJS..171..260L, 2009AJ....138..312H}.  So that the correlation between Ca~II and photometry displayed by the Sun becomes an anticorrelation for more active stars.  Thirdly,  investigations comparing  the Sun with other main-sequence stars \citep{2007ApJS..171..260L, 2013ASPC..472..203L} showed that solar brightness variability over the 11-year activity cycle appears to be anomalously low compared with stars of near-solar magnetic activity. \\
The latter observation was  used to suggest that historical solar variability and consequently the solar role in pre-industrial climate change might have been significantly larger than thought before \citep[see, e.g.~][]{doi:10.1029/92GL01578,  2011A&A...529A..67S, 2012A&A...544A..88J, 2013ARA&A..51..311S}. Another recently proposed explanation \citep{2016A&A...589A..46S} is based on the fact that solar brightness variability is caused by a delicate balance between dark and bright magnetic features. This balance is sensitive to the combination of stellar fundamental parameters, which define properties of magnetic features, i.e.~effective temperature, metallicity and surface gravity. Consequently,  stars with slightly different fundamental parameters can show significantly altered, e.g.~higher, brightness variations.  So far this hypothesis could not be tested due to the absence of reliable stellar atmospheric models for magnetic features, which hindered the development of quantitative models of stellar brightness variability for a long time. \\
In contrast to other stars, accurate model atmospheres of quiet regions and magnetic features exist for the Sun.  The variability of Sun-like stars can reasonably be assumed to be based on the same mechanisms as on the Sun, where the  processes that are responsible can be observed in detail. Therefore,  it is possible to extend solar irradiance models \citep[see][and references therein]{acp-13-3945-2013, 2013ARA&A..51..311S} 
to investigate Sun-like stars.  Previous studies suggested several approaches depending on the issue of interest. For example,  the effect of inclination for stars observed out of their equatorial plane on the photometric variability \citep{1993JGR....9818907S, 2001A&A...376.1080K, 2012GeoRL..3916104V, 2014A&A...569A..38S} was studied and a potential  increase of variability  was found for large inclinations.    \citet{2009A&A...493..193L} developed different brightness variability models that account for active regions, but which have at least eleven free parameters, to analyse and fit observed light curves. 
Moreover, the correlation between the faculae and spot-dominated stellar variability and magnetic activity was investigated by modelling a hypothetical Sun at different activity levels and  extrapolating the  surface coverage  by solar magnetic features to another mean chromospheric activity level \citep{2014A&A...569A..38S}.
 Whereas these investigations can explain the  activity dependence  of variability as well as the transition from faculae- to spot-dominated stars, they did not address the anomalously low brightness variability of the Sun, which remains a long-standing puzzle. \\
%
%
%
Generally, stars are characterised as Sun-like or even solar twins if their fundamental stellar parameters are close to the solar parameters. While such stars are close to the solar case, no star has the identical set of fundamental stellar parameters. Here, the goal is to shed light on the role of several stellar fundamental parameters for stellar brightness change, in particular different metallicities and effective temperatures.
For that, we extend the successful brightness variations model SATIRE  \citep{2000A&A...353..380F, 2003A&A...399L...1K}  to Sun-like stars with different fundamental stellar parameters. Thus we take advantage of building on the successful existing solar models that have been developed for decades and agree accurately with the solar observations. Whereas most stellar brightness variation studies  assumed solar contrast of magnetic features, we take a different path. Our novel approach is based on calculating the spectral contrasts of spot umbrae, spot penumbrae, and faculae to the quiet stellar region depending on stellar fundamental parameters. This allows us to investigate the influence of stellar fundamental parameter, e.g.~metallicity and effective temperature,  on stellar brightness variations. Here, we limit our study to stars with the same surface distribution of magnetic features as the solar case.  The main goal is to  investigate stellar brightness change, in particular in the Str\"omgren {\it{b}} and {\it{y}} filters and the Kepler passband, on the time-scale of magnetic activity cycles. \\
%
%
%
 This paper is structured as follows. In Section~\ref{sec:theoretical_approach} our theoretical approach is described. In Section~\ref{sec:result} we present our  results of the effect of the metallicity, inclination, and effective temperature on  stellar brightness change. Finally, we provide a summary and draw conclusions in Section~\ref{sec:conclusion}.  
\section{Model: From the Sun to Stars} 
\label{sec:theoretical_approach}

\begin{figure*}
  \centering
   \includegraphics[width=0.45\linewidth]{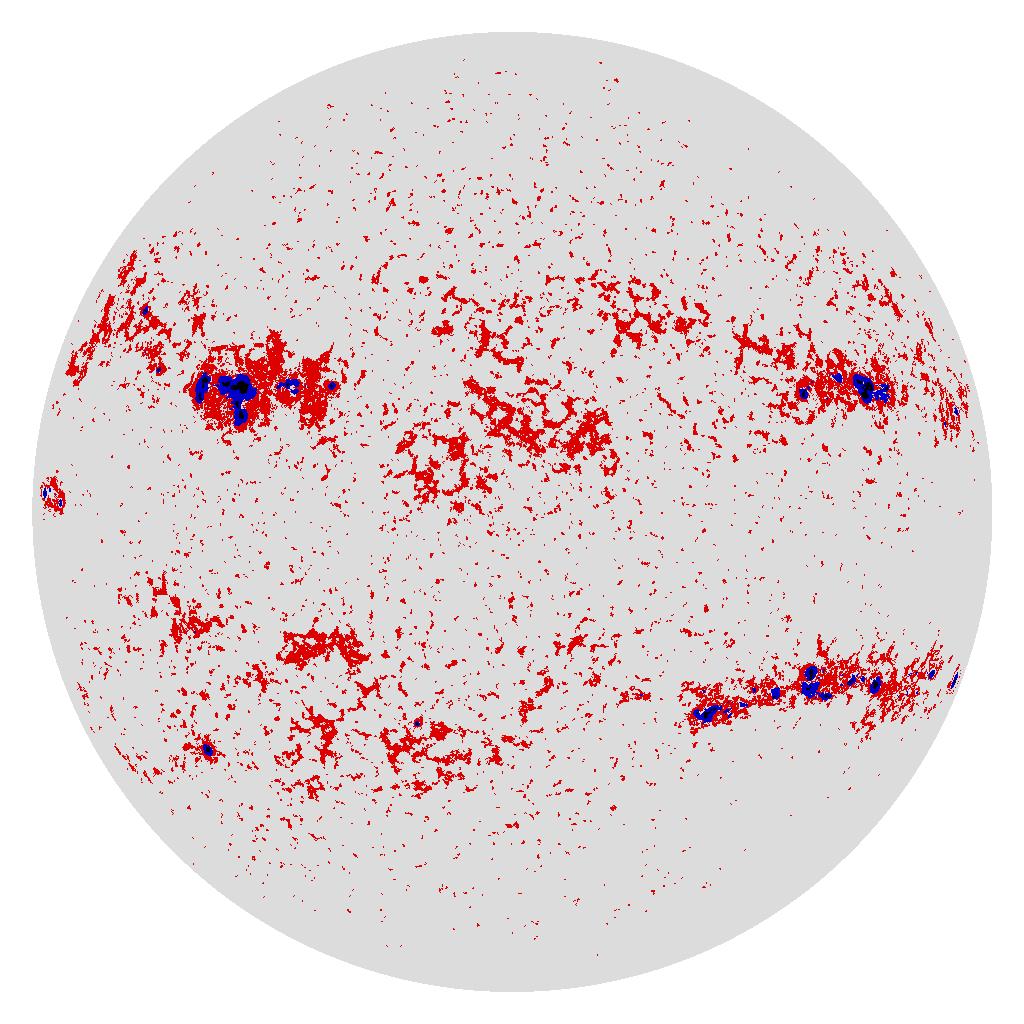}
    \includegraphics[width=0.45\linewidth]{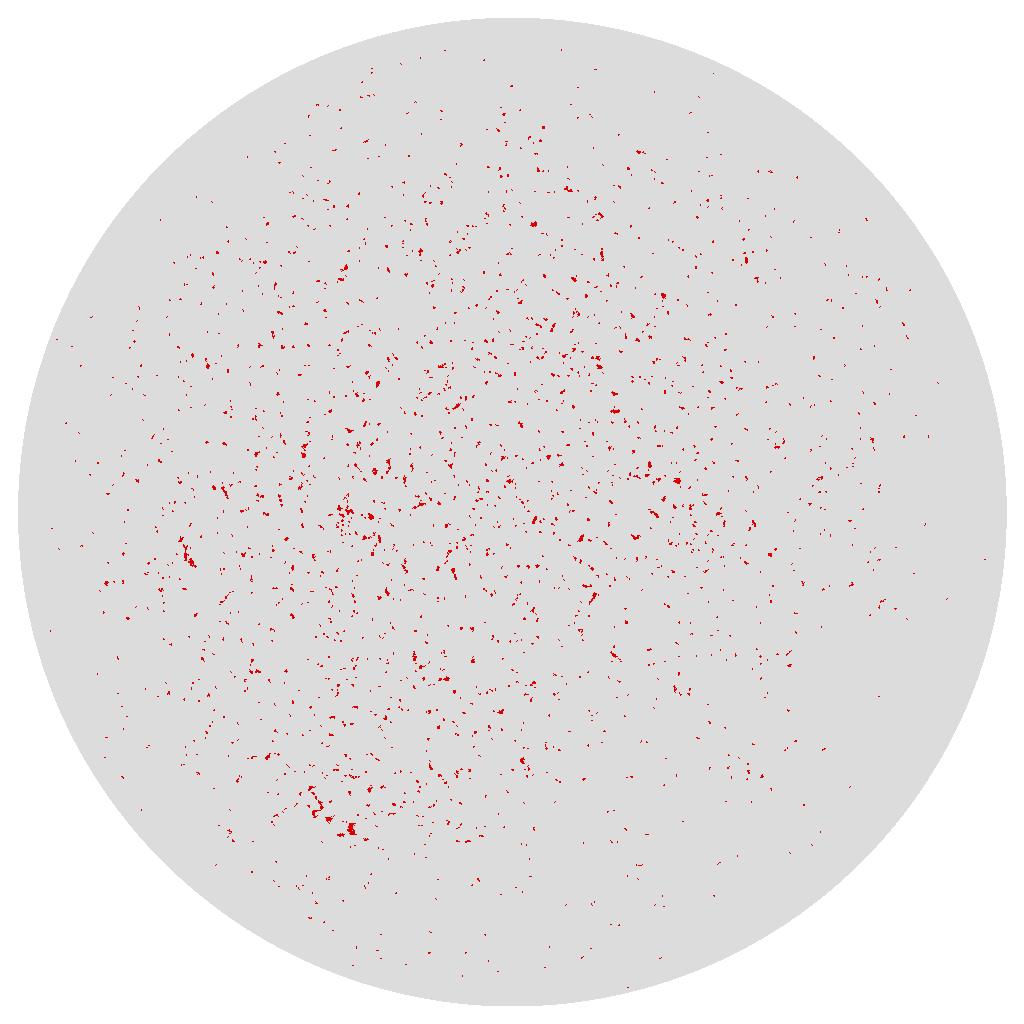}
   \includegraphics[width=0.80\linewidth]{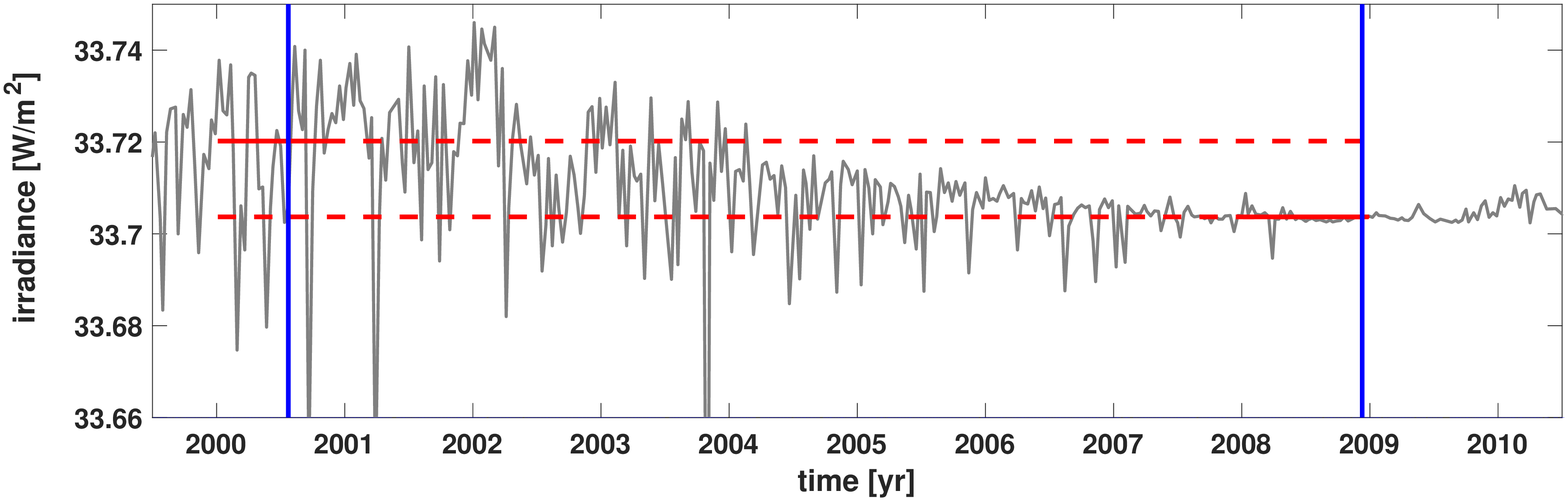}
      \caption{Distribution  of magnetic features on the solar surface (upper row) and computed irradiance variations (lower panel). Upper plots:   Black areas represent umbra, blue regions denote penumbra and red areas are faculae. Top left: 11th of July, 2000. Top right: 1st of Dec, 2008.  Lower plot: Solar irradiance in Str\"omgren \textit{b} filter. Here, we normalise the total amount of spectral energy to one AU. The vertical blue lines represent the times of the two upper plots. The two horizontal lines represent the average irradiance levels in the years 2000 and 2008. }
         \label{fig:model00}
   \end{figure*}

In this study we focus on stellar photometric brightness change on the magnetic activity cycle time-scale,  which for the Sun is approximately 11 years. Note, another important time-scale is the rotational time-scale, which will be investigated separately in a follow-up paper. To obtain a time dependent spectrum the observed magnetic feature distributions of solar cycle 23 are used, which was a cycle of intermediate strength. The effect of the cycle strength on the brightness variations was investigated in \citet{2014A&A...569A..38S}, and here we choose a representative cycle for which we have good measurements of solar surface magnetic field. The typical distributions of faculae and spots on the solar surface during the maxima and minima of the solar cycle are displayed in Fig.~\ref{fig:model00}, together with the time-evolution of the solar irradiance in Strömgren filter {\it{b}}.  The difference between the year 2000 and 2008, which is indicated by a red horizontal lines in the bottom panel of Fig.~\ref{fig:model00},  is taken here to represent the brightness change on the activity cycle time-scale.  \\
\subsection{Photometric Brightness Change}
The SATIRE model that we describe here is used to compute the photometric brightness change. The SATIRE model separately accounts for  the quiet stellar regions, star spot umbral, star spot penumbral and facular components \citep{2003A&A...399L...1K}. Following the detailed description in \citet{2014A&A...569A..38S} the spectral flux can be decomposed into two main contributions: 
\begin{equation}
\label{eq:total_flux}
F(\lambda) = F_{Q}(\lambda) + F_{m}(\lambda),
\end{equation}
where $Q$ denotes the quiet part of the stellar surface, $m$ is associated with different magnetic features and $\lambda$ is the wavelength.  Then, the disk integrated flux $F_Q(\lambda)$ is obtained by integrating  
\begin{equation}
F_Q (\lambda) = \int_0^1 I_Q(\lambda, \mu) \omega(\mu) d\mu,
\end{equation}
where  $\omega(\mu) = 2\pi \mu (r_{star}/ d_{star})^2$ is a weighting function with the stellar radius, $r_{star}$, and the distance between the star and observer, $d_{star}$. The considered emergent intensity, $I_Q(\lambda, \mu )$,  depends also on $\mu$, which is the cosine of the angle between the observer's direction and the local stellar radius.  The magnetic features contribute through  their contrast $I_{m}(\lambda, \mu) - I_Q(\lambda, \mu)$ to the stellar  brightness
\begin{equation}
\label{eq:magnetic_features_flux}
F_{m}(\lambda )= \int_0^1 \sum_{m} \alpha_{m}(\mu) \left(I_m(\lambda, \mu) - I_Q (\lambda, \mu) \right)\omega(\mu) d\mu,
\end{equation}
where the fractional coverage of the ring at the corresponding $\mu$  on the stellar disk is given by the functions $\alpha_{n}(\mu)$. In this formulation the coverage at the stellar disk center is associated with $\alpha_{m}(\mu =1)$ and at the limb with $\alpha_{m}(\mu =0)$.  In order to gain a more detailed understanding of separate contributions from faculae and spots, $F_{m}$ can be further decomposed. The facular spectral flux is defined as
\begin{equation}
\label{eq:Faculae_flux}
F_{Fac}(\lambda )= \int_0^1  \alpha_{Fac}(\mu) \left(I_{Fac}(\lambda, \mu) - I_Q (\lambda, \mu) \right)\omega(\mu) d\mu,
\end{equation} 
whereas the spectral flux from spots consists of both the spot umbra and spot penumbra components:
\begin{eqnarray}
\label{eq:Spot_flux}
F_{spot}(\lambda )&=&    \int_0^1 \alpha_{Pen}(\mu) \left(I_{Pen}(\lambda, \mu) - I_Q (\lambda, \mu) \right)  \omega(\mu) d\mu   \nonumber \\
 &+ & \int_0^1    \alpha_{Umb}(\mu) \left(I_{Umb}(\lambda, \mu) - I_Q (\lambda, \mu) \right)   \omega(\mu) d\mu.
\end{eqnarray}
The solar intensities, $I(\lambda, \mu)$,  and solar surface coverages, $\alpha_m(\mu)$, fully determine the solar Sun.  In previous investigations the magnetic feature coverages, $\alpha_m(\mu)$, were varied to model different stars \citep{2001A&A...376.1080K, 2009A&A...493..193L, 2014A&A...569A..38S}, whereas the emergent intensities, $I_m(\lambda, \mu)$ and $I_Q(\lambda, \mu )$, and thus the contrasts of the magnetic features were fixed to be as on the Sun. Since the aim of this investigation is to understand potential differences between brightness changes on a hypothetical Sun with slightly different fundamental parameters, we utilize a complementary approach and  use the same coverages of stellar features as for the solar case, but contrasts of the  magnetic features and their  center-to-limb variations (CLVs) will be adopted by recalculating the emergent intensities from the quiet and magnetic stellar regions.\\
\subsection{Radiative Transfer Model}
\label{Subsec:Radiative_transfer_m}
 \begin{figure}
   \centering
   {\includegraphics[width=1\linewidth]{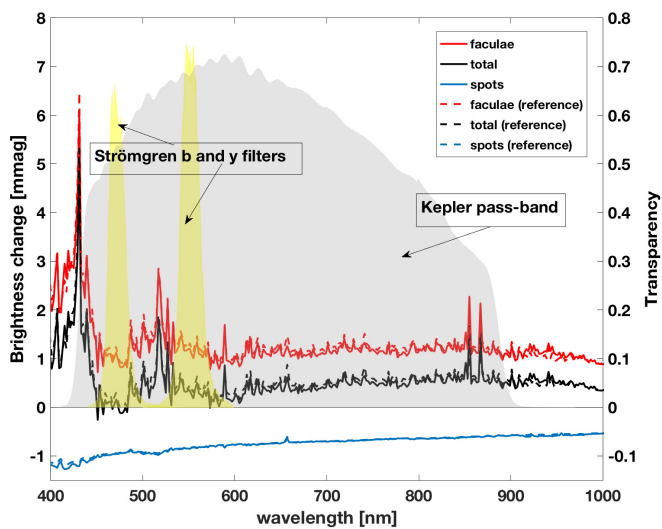}}
      \caption{The amplitude of solar cycle brightness change, here the difference between annually averaged solar brightness in 2000 (cycle max) and 2008 (cycle min), calculated with the SATIRE model (black). In addition, we plot the facular and spot components of the brightness change (red and blue). The gray shaded area indicates the spectral response function of the \textit{Kepler} telescope measurements and the yellow shaded areas the Str\"omgren filters \textit{b} and \textit{y} (centered at 467~nm and 547~nm, respectively). Solid lines: Solar cycle brightness change calculated with the newest ATLAS9 code and ODFs. Dotted lines: Same but using emergent spectra calculated by \citet{1999A&A...345..635U}. }
         \label{fig:model02}
   \end{figure}

In this study we are interested in the amplitude of solar cycle brightness change, defined here as the difference between annually averaged solar brightness in 2000 (cycle max) and 2008 (cycle min). To obtain the brightness change using the Equations~\eqref{eq:total_flux}~-~\eqref{eq:magnetic_features_flux},   the emergent spectra for the different stellar components are needed. Those can be calculated by solving the radiative transport equation for the corresponding atmospheric models. For the photospheric layers of  the quiet Sun, spot umbra, and spot penumbra such atmospheric models are often computed assuming radiative equilibrium while convection is included through mixing length theory \citep{1958ZA.....46..108B}. The faculae models are semi-empirical  \citep{1981ApJS...45..635V, 1981A&A...103..160L, 1993ApJ...406..319F, 2006ApJ...639..441F}, i.e. they are constructed to match their output to solar observations recorded at intermediate spatial resolution \citep{1981ApJS...45..635V}.  Thus, faculae models are not in radiative equilibrium.  Despite the fact that these models do not account for the geometric properties of magnetic features such as hot walls in magnetic flux tubes \citep{1993SSRv...63....1S, 2005A&A...430..691S}, ensembles of which form network and faculae,  they have been successfully used for many applications, in particular, for modelling solar brightness variations \citep{2013ARA&A..51..311S}.   \\
In order to obtain the emergent intensities across a wide range of wavelengths from the atmospheric models, we use the LTE spectral synthesis code ATLAS9 \citep{1992RMxAA..23...45K, 1994A&A...281..817C}. For the calculation of the stellar continuum opacities, the following contributors are taken into account: Free-free (ff) and bound-free (bf) transitions in $\text{H}^-$, $\text{H I}$, $\text{H}_2^+$, $\text{He I}$,  $\text{He II}$, $\text{He}^-$. In addition, ff and bf transitions for low to high temperature absorbers such as C, N, O, Ne, Mg, Al and Si.  Moreover, electron scattering and Rayleigh scattering on $\text{H I}$, $\text{He I}$ and $\text{H}_2$ are considered. ATLAS9 further exploits Opacity Distribution functions (ODFs) to account for the opacity of  millions of atomic and molecular spectral lines. For that the ODFs are generated by the code DFSYNTHE~\citep{2005MSAIS...8...34C} for two microturbulence velocities of $1.5 \text{km s}^{-1}$ for the quiet stellar component and facular component, and of $2.0 \text{km s}^{-1}$ for the spot components, where a higher velocity is chosen to partly account for the large Zeeman splitting in the spots \citep{0004-637X-639-1-441, Anderson_2010}.  To account for the center-to-limb variation, the emergent intensities, $I(\lambda, \mu)$,  are calculated for 11 different $\mu$ values. \\
So far  the SATIRE model made mainly use of pre-calculated emergent spectra by \citet{1999A&A...345..635U} to successfully reconstruct the solar irradiance variations.  For our purpose it is necessary to calculate these emergent spectra for different fundamental stellar parameters. The ATLAS9 code has evolved since \citet{1999A&A...345..635U} used it to compute the emergent spectra generally needed for the SATIRE model. Therefore, we recomputed the change in the solar spectrum between 2000 \& 2008, as well as the facular and spot contributions to this change (shown in Fig.~\ref{fig:model02}) by exploiting the newest ATLAS9 version (together with recalculated ODFs for solar elemental composition taken from  \citep{1989GeCoA..53..197A}).  For that we use the same atmospheric structures as \citet{1999A&A...345..635U} for the quiet sun, sun spots, and faculae as our reference. The so obtained brightness changes agree very well with previous calculations by the SATIRE model based on the \citet{1999A&A...345..635U}  spectra (see dotted lines in Fig.~\ref{fig:model02}). The small deviations are  due to up-dates incorporated over the last decades in the ATLAS9 and DFSYNTHE~codes.\\
%
%
%
Now that we have validated the approach for the solar case, let us consider different metallicity values. These affect the spectrum in two ways. Firstly, they change the strengths of atomic and molecular lines. Secondly, for a cool star, changed metallicity values influence the continuum opacity, due to the change in the concentration of electron donors. We consider these effects separately, to judge the importance of each. 
In a first step, only the direct effect of the metallicity on the atomic and molecular lines is considered on its own. For this, we use the corresponding recalculated ODFs for the new metallicity, while keeping the atmosphere models used by \cite{1999A&A...345..635U} for the quiet Sun and magnetic features, thus neglecting the effect of a different metallicity on the atmosphere's structure and the electron concentration. The results of this approach are presented in Subsection~\ref{subsec:Fraunhofer_lines}. In the next step, both, the effect of lines as well as the change of the electron concentration along with the back-reaction of the changed radiation field on the atmospheric structure is considered. For that the ATLAS9 code provides a tool, which self-consistently calculates 1D radiative equilibrium  atmosphere models for different fundamental stellar parameters, i.e.~effective temperature and metallicity.  However, since the facular model was obtained semi-empirically, and no three-dimensional magnetohydrodynamic calculations (3D MHD)  based calculations of magnetic features  for different metallicities are currently available, the only option is to modify the reference facular model. The empirical modification of the atmospheric model for the faculae and the results of this approach are discussed in Subsection~\ref{subsec:adjust_atmo_models}.\\
%
%
%
\section{Results}
\label{sec:result}
To understand how fundamental stellar parameters affect the brightness change of a hypothetical Sun, we vary the metallicity, the inclination, and the effective temperature separately. 
We begin our study by investigating the effect of metallicity on stellar brightness change.  For that we first consider only  the effect  of  Fraunhofer lines on the opacity (Subsection~\ref{subsec:Fraunhofer_lines}), where at this first stage we neglect the feedback on the atmospheric structure. We then analyse the combined effect due to Fraunhofer lines, changed electron number density and changed stellar atmospheric structure (Subsection~\ref{subsec:adjust_atmo_models}). The effect of inclination and effective temperature is considered in Subsections~\ref{:subsection:inclination} and \ref{subsec:effective_temp}, respectively. 

\subsection{Direct Effect of Metallicity  on  Fraunhofer Lines}
\label{subsec:Fraunhofer_lines}
 \begin{figure}
   \centering
   {\includegraphics[width=1.0\linewidth]{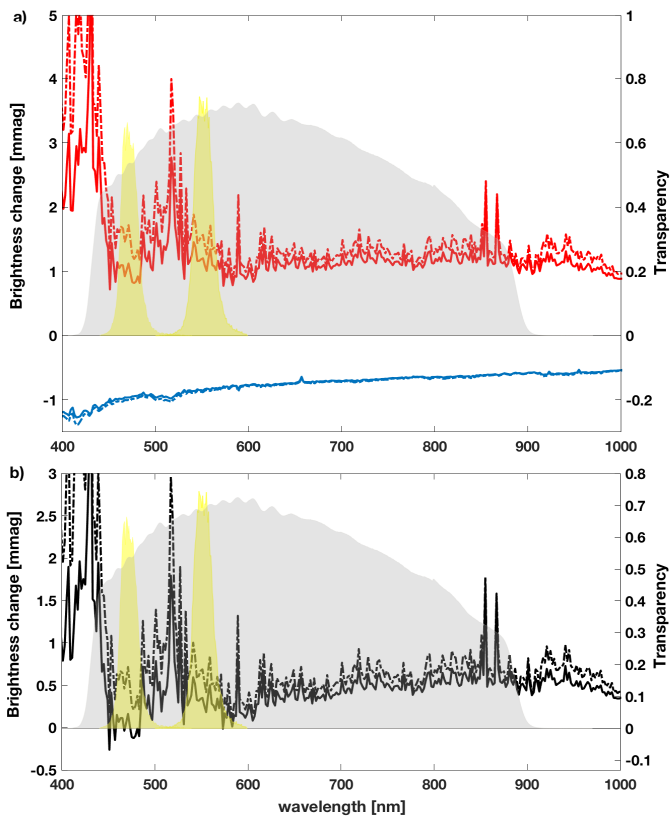}}
      \caption{a) Radiative flux difference between 2000 and 2008 for different atmospheric components and two metallicity values M/H.  Facular (red) and spot (blue)  contributions for M/H values of 0.0 (solid lines) and 0.2 (dashed lines). b) Same as in panel a), but now for the facular and spot contributions combined. The yellow and grey shaded areas are the same as in Fig.~\ref{fig:model02}.  }
        
         \label{fig:Fraunhofer01}
   \end{figure}
 \begin{figure}
   \centering
   {\includegraphics[width=1.0\linewidth]{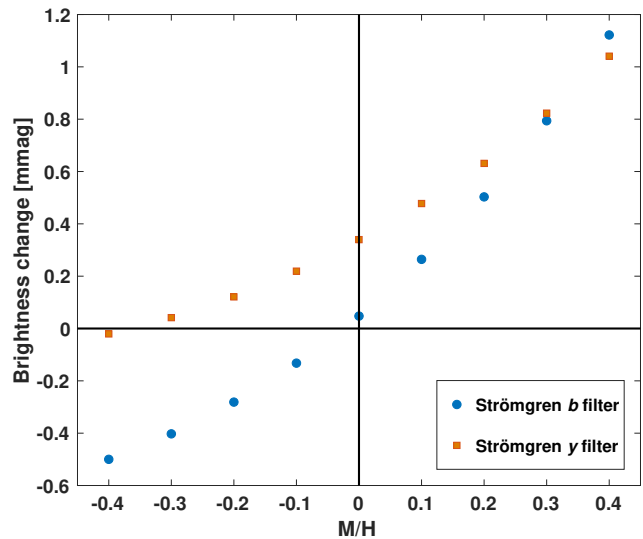}}
      \caption{Brightness change integrated over Str\"omgren \textit{b} and \textit{y} filters as a function of metallicity}
         \label{fig:Fraunhofer02}
   \end{figure}
On time-scales greater than a day,  brightness variations are determined by changes in the surface-area coverage by magnetic features  \citep[e.g.][]{2013ARA&A..51..311S, 2017PhRvL.119i1102Y}. Therefore, the sum of the facular and star spot contributions  is decisive for the amplitude of the  photometric brightness change. Furthermore, \citet{2015A&A...581A.116S} showed that for the Sun, which is faculae-dominated,  the main contribution to the brightness change on solar cycle time-scales in the UV and visible spectral domains comes from the molecular and atomic lines that are present in the emergent spectra of faculae \citep[see also][]{1991ApJ...372..336M, Unruh2000}. 
This has to do with the particular temperature structure of faculae.  While their temperature is that of the quiet Sun near the Rosseland mean optical depth,  $\tau_{Ross}$, equal to unity, they are considerably hotter in the higher layers. Hence, at the disc centre their continuum contrast is relatively low, while in the cores of most spectral lines of neutral elements they are bright \citep{2013A&A...550A..95Y}. This is partly because, the spectral lines of neutral atoms and molecules are weakened in faculae due to enhanced ionization and dissociation, respectively. \\
The strengths of the lines (and especially of the weak unsaturated lines) depend on the
metallicity, so that any change of the metallicity should affect the amplitude of the stellar brightness change. To quantify this effect we first compute emergent spectra by using ODFs for the relevant metallicities. \\
We calculate a set of ODFs with different metallicities, which are defined as
\begin{equation}
\text{M/H} = \log{\left( \frac{\left( N_{\text{metals}} / N_{\text{H}} \right)_{\text{star}} }{ \left( N_{\rm metals} / N_{\rm H} \right)_{\rm Sun}} \right) },
\end{equation}
where $N_{\rm metalls}$ is the number density of all metals and $N_{\rm H}$ the number density of hydrogen. We start with the solar metallicity, $\text{M/H} = 0.0$, and the element composition by \cite{1989GeCoA..53..197A}. To obtain metallicities in the range ${-0.4 \le \text{M/H} \le 0.4}$ in steps of $0.1$, we scale the ratio of all metalls to hydrogen accordingly. The ODFs are then used together with the reference atmospheric models to calculate the emergent intensities for  Equations~\eqref{eq:total_flux}~-~\eqref{eq:Spot_flux} (SATIRE model). At this step, the approach is, however, not yet self-consistent. While the ratio of all metals to hydrogen is changed and thus the formation height of different lines is shifted,  the $H^{-}$ opacity, which is the main contributor to the continuum in the visible, remains unchanged because neither the electron density nor the temperature structure is re-calculated. Note that in this approach, when only the metallicity for the line opacities is changed the effective temperature, $T_{\rm eff}$, is still somewhat affected, e.g.~for higher metallicities $T_{\rm eff}$ becomes lower due to stronger lines.  \\
The brightness change computed in this way is shown in Fig.~\ref{fig:Fraunhofer01}.  Figure~\ref{fig:Fraunhofer01} a) shows the effect of an increased metallicity $\text{M/H} = 0.2$  for the facular  and  spot contributions separately,  while Fig.~\ref{fig:Fraunhofer01} b) shows the effect on the overall brightness change. The solar case as  in Fig.~\ref{fig:model02} is also plotted for comparison. As explained earlier in this section, the main reason why the Sun is brighter during activity cycle maximum are the weaker spectral lines in the faculae \citep{2015A&A...581A.116S}. If we enhance metallicity, we increase the strength of spectral lines, so that to first order more lines get weakened in faculae and their contrast increases. Because the line density is higher in the UV, the increase in contrast is also larger there, as confirmed by Fig.~\ref{fig:Fraunhofer01} b).
In contrast, the spot contribution changes little (Fig~\ref{fig:Fraunhofer01} a). This can be understood by looking at the spectral profiles of the faculae and spot contributions to the brightness change: While the facular profile contains a lot of spectral features brought about by the Fraunhofer lines, the spot profile is pretty smooth.  This indicates that the contrast due to Fraunhofer lines is predominant for the faculae component, while continuum plays a more important role for the spot component. This is supported by the temperature profile of sunspot umbrae and penumbra, which have similar (or flatter) gradients than the quiet Sun, but a significantly lower effective temperature. \\ 
Stellar photometric brightness variation measurements spanning a decade or more have pre-dominantly been made in the Str\"omgren \textit{b} and \textit{y} filters. Thus the dependence of brightness change on metallicity in these filters, shown in Fig.~\ref{fig:Fraunhofer02}, is of particular interest. The brightness change in these filters is also particularly sensitive to metallicity. This is partly because facular and sunspot variations nearly balance each other for solar metallicity, especially in Str\"omgren $b$.  For $\text{M/H} = 0.2$, for example, the brightness change in the Str\"omgren filter \textit{b} is increased by a factor of $10.6$ relative to the Sun, whereas in the Str\"omgren filter \textit{y} it is increased by a  smaller, but still sizeable factor of $1.86$.  \\
Interestingly, for metallicities smaller than in the Sun, the brightness change becomes negative in the Str\"omgren \textit{b} filter, which indicates that for low metallicities the brightness changes over the  activity cycle are spot-dominated rather than faculae-dominated. Both regimes have been observed in main-sequence stars \citep{1998ApJS..118..239R,2007ApJS..171..260L, 0004-637X-851-2-116}, where magnetically less active and consequently older stars usually show a direct correlation between brightness and activity, i.e. they are faculae-dominated, while more active stars are spot-dominated. It was observationally established that the transition between these two regimes is approximately at a chormospheric activity of $\text{log R'}_{HK} \approx  -4.7$,  although the two regimes overlap \citep{0004-637X-851-2-116}. Our finding that a change in metallicity can lead to a change from faculae- to spot-dominated brightness variations for the same activity level is one possible explanation for this observation. 
\subsection{Adjusting Atmospheric Models to Changed Metallicity}
\label{subsec:adjust_atmo_models}
\begin{figure}
   \centering
   {\includegraphics[width=1.0\linewidth]{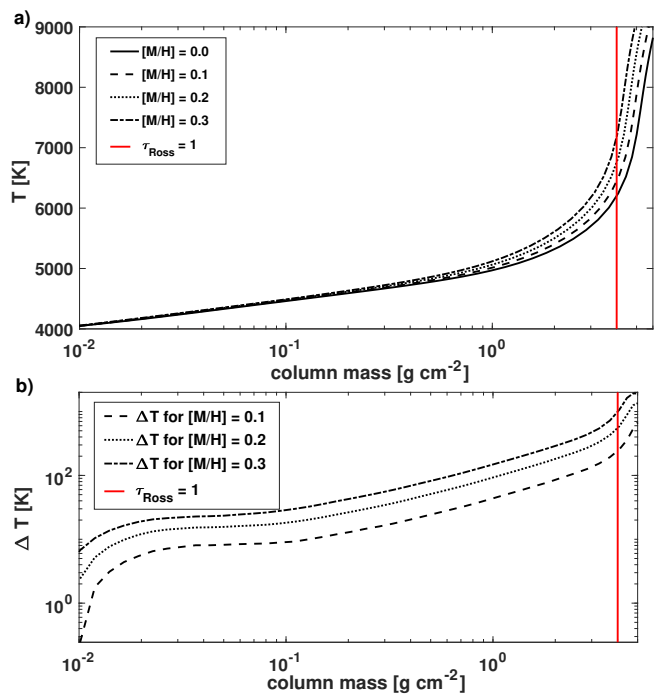}}
      \caption{Self-consistently calculated atmospheric models for different metallicities. All models are constructed for  $T_{\rm eff} $ of the quiet sun model, which is $T_{\rm eff} = 5777$K. a) Temperature structure of quiet stellar region, vs. column mass. b) The temperature change,  $\Delta T$,  with respect to the solar metallicity value $\text{M/H} = 0.0$ with column mass.  The red line indicates $\tau_{Ross} = 1$ layer for the quiet Sun's (M/H~=~0.0) atmosphere.}
         \label{fig:Adjust01}
   \end{figure}
\begin{figure}
   \centering
   {\includegraphics[width=1.0\linewidth]{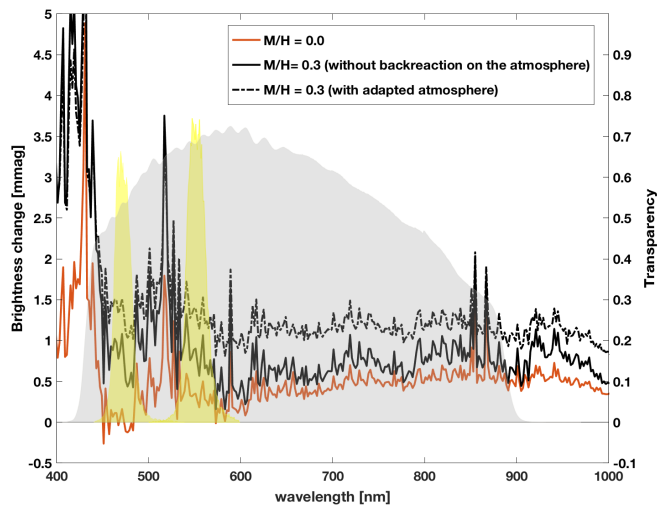}}
      \caption{The amplitude of solar cycle brightness change as defined in Fig.~\ref{fig:model02}. Orange solid line: the solar case (M/H~=~0.0).  Black solid line: A hypothetical Sun calculated with M/H~=~0.3 using solar atmosphere models. Black dotted line: same as before, but using atmospheric models consistently recalculated with M/H = 0.3, i.e. the back-reaction of the radiation field on the atmospheric structure, and continuum opacity change due to a different electron concentration are considered. The yellow and grey shaded areas are the same as in Fig.~\ref{fig:model02}.}
         \label{fig:Fig01}
   \end{figure}
\begin{figure}
   \centering
   {\includegraphics[width=1.0\linewidth]{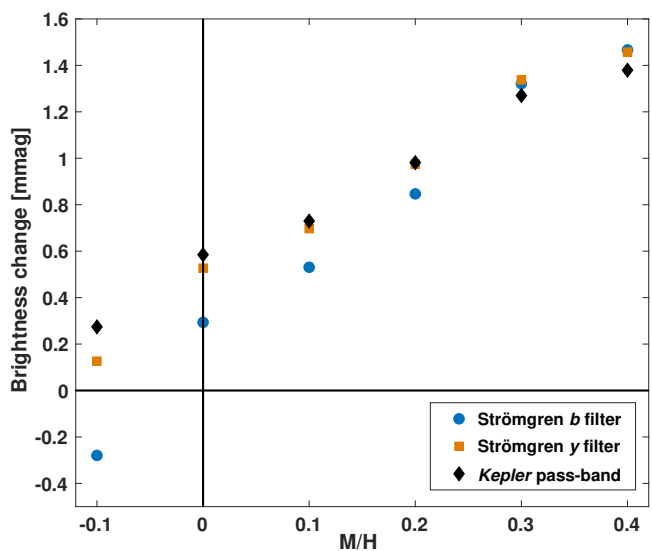}}
      \caption{Change in radiative flux integrated over Str\"omgren \textit{b} and \textit{y} filters and \textit{Kepler} pass-band as a function of metallicity for cases with recalculated atmosphere models.}
         \label{fig:Adjust03}
   \end{figure}

While in Section~\ref{subsec:Fraunhofer_lines} we focused on the effect of metallicity directly on the opacity of atomic and molecular lines, we did not take into account the change in the electron concentration that affects the $\text{H}^{-}$ concentration and consequently the continuum opacities. Recalculating the overall electron concentration leads to a significant change in the radiation field, and a shift of the continuum formation height, which results in a different effective temperature. In order to restore the initial  effective temperature, we account for  the changed radiation field by recalculating the atmospheric structure to be in radiative equilibrium. 
In this section the aim is to investigate all effects together,  the Fraunhofer lines, recalculated electron concentrations, and the changed atmospheric structure for different metallicities in the range $-0.1 \leq \text{M/H} \leq 0.4 $.\\
In order to quantify the full effect of metallicity on the brightness change of a star, it is therefore necessary to  re-calculate models for the corresponding M/H value as described in Subsection~\ref{Subsec:Radiative_transfer_m}. Models of the quiet stellar regions, umbra, and penumbra can be self-consistently calculated with ATLAS9 assuming radiative equilibrium and preserving their effective temperatures for a prescribed M/H value. In Fig.~\ref{fig:Adjust01} a) the re-calculated atmospheric structures for the quiet stellar regions are displayed for four different M/H values, whereas the resulting $\Delta \text{T}$ between a particular metallicity value and the solar case ($\text{M/H} = 0.0$) is shown in Fig.~\ref{fig:Adjust01}~b). Focusing on the location of the Rosseland optical depth, $\tau_{Ross} = 1$, in this plot, shows that the greatest atmospheric changes of a few hundred K are at depths where the continuum is formed. This can be explained as follows. On the one hand, for increased metallicity spectral lines become stronger, the emerging radiation decreases, and so does $T_{\rm eff}$. To compensate for this decrease in emerging radiation, and return to the initial $T_{\rm eff}$, in a self-consistent atmosphere the temperature of the continuum-forming layer has to be increased. The stronger continuum offsets the deeper spectral lines, so that $T_{\rm eff}$ remains unchanged.  On the other hand, increased opacity due to higher metallicity leads to a slight shift of the  $\tau_{Ross} = 1$ on the column mass scale, so that the continuum is formed at somewhat higher layers compared to the solar case.\\ 
As the facular model is not in radiative equilibrium, it cannot be directly re-calculated by the ATLAS9 code. Therefore we assumed that a changed  metallicity value has the same effect on the temperature structures of the  faculae as on the quiet stellar regions, applying the $\Delta \text{T}$ shown in Fig.~\ref{fig:Adjust01}~b) to the solar facular model. In other words, we assumed that the change of the metallicity preserves the temperature contrast between the faculae and  the quiet regions as a function of the column mass. We plan to test this assumption in future with a realistic 3D MHD calculation, which, however, is beyond the scope of the present paper.\\
Taking all effects combined into account has a large impact on stellar photometric brightness change as shown by the difference between the solid black line and the dashed black line in Fig.~\ref{fig:Fig01}. In this particular case, for a metallicity value of $\text{M/H} = 0.3$, the radiative flux difference between activity maximum and minimum in the Str\"omgren \textit{b} and \textit{y} filters is approximately 1.5 times greater when we consider recalculated and adjusted atmospheric models compared to the case considered in the previous Subsection~\ref{subsec:Fraunhofer_lines}. \\
Finally, the effect of metallicity on the brightness changes in the Str\"omgren \textit{b} and \textit{y} filters, and the {\it{Kepler}} pass-band  is shown in Fig.~\ref{fig:Adjust03}. In this set of  calculations the combined effect of lines, recalculated atmospheric model and changed electron concentration is captured.  Note, that the solar case is not the same reference as in the previous section. This is because \citet{1999A&A...345..635U} adjusted  the temperature structure of the initial facular model for the solar case to match the observed emergent spectra without changing the electron concentration.  Recalculating the electron concentration for the adjusted temperatures in the facular model leads to a greater brightness change and consequently a shifted reference point. The combined effect leads to an even stronger increase in brightness change with greater metallicity than was found  in Subsection~\ref{subsec:Fraunhofer_lines} for the isolated effect of metallicity on Fraunhofer lines. A transition to a spot-dominated regime is found for the Strömgren \textit{b} filter close to $\text{M/H}= -0.1$, where a similar result was found in Subsection~\ref{subsec:Fraunhofer_lines}. \\
These findings support the hypothesis that the balance between radiation from dark and bright magnetic features that determines the brightness change on the magnetic activity cycle  time-scale is sensitive to stellar fundamental parameters. In particular, the solar metallicity value is close to a complete compensation regime, which is one possible explanation for the low brightness change of the Sun compared to other Sun-like stars of similar chromospheric activity \citep{2007ApJS..171..260L, 2013ASPC..472..203L}.  Previous analysis of the relationship between chromospheric activity measured in $\text{log} R'_{HK}$  and photometric variations in Str\"omgren $(b + y )/2$  for Sun-like stars revealed a power-law relation \citep{1992Natur.360..653L, 2007ApJS..171..260L}.  The Sun, however, was found to be located significantly below the established power-law relation \citep{ 2007ApJS..171..260L,  2014A&A...569A..38S, 2018ApJ...855...75R}. According to our calculations a hypothetical Sun with a metallicity of  $\text{M/H} = 0.4$ would have a photometric variations for $(b + y )/2$ of $\text{log rms} = 0.52 \times 10^{-3} $ and thus it would lie close to the power-law relation found by  \citet[][their Figure 7]{2007ApJS..171..260L}. 
\subsection{Inclination Effects}
\label{:subsection:inclination}

\begin{figure}
   \centering
   {\includegraphics[width=1.0\linewidth]{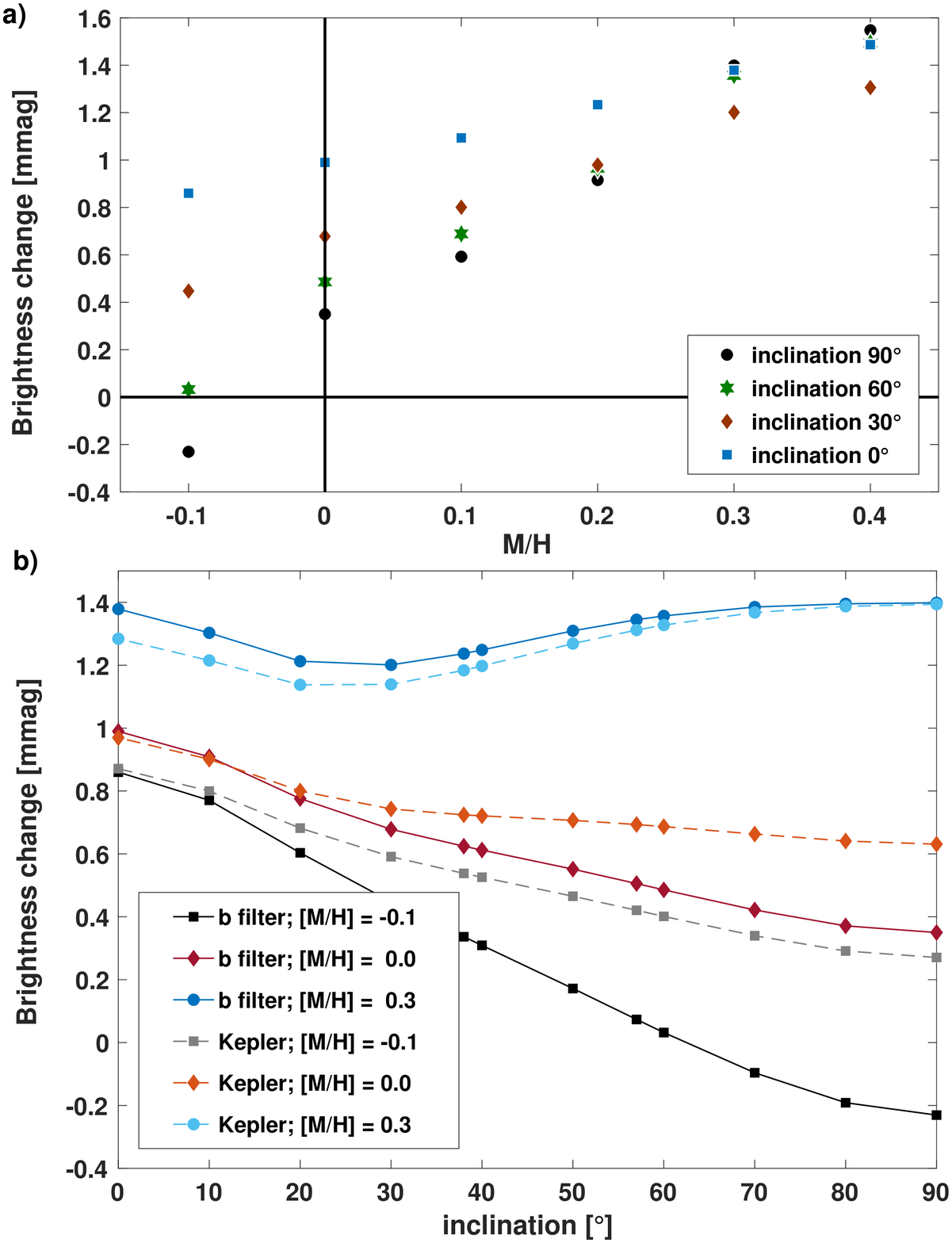}}
      \caption{Brightness change in the Strömgren \textit{b} and Kepler  filters for different inclination angles ($90^{\circ}$ corresponds to an equator-on view) and different metallicity values. a) Brightness change vs. metallicity for four different inclination angles. b) Brightness change vs. inclination angle, for three different metallicities and two different filters (Str\"omgren $b$ and Kepler). }
         \label{fig:inclination01}
   \end{figure}

The Sun  is always  observed from a close to an equator-on vantage point.  This is different for other stars, which are  observed at random inclination angles, defined as the angle between the observer's viewing direction  and the rotational axis of the star.  We define the inclination angle such that a pole-on view corresponds to $0^{\circ}$ and  an equator-on view to  $90^{\circ}$.  Several investigations studied  the effect of inclination on the solar brightness variation on the magnetic activity cycle time-scale \citep[e.g.,][]{1993JGR....9818907S, 2001A&A...376.1080K, 2014A&A...569A..38S}. All previous studies found that an inclined Sun would show a greater brightness variation, although they differed in the magnitude of the effect of inclination. However,  prior studies considered only the solar case, and did not consider other stellar fundamental parameters. 
Here, we investigate the effect of the inclination for stars with different metallicities, with the focus on the brightness change in the Strömgren \textit{b} and Kepler filters. Using the calculated spectra for different metallicities with the recalculated model (see Section~\ref{subsec:adjust_atmo_models}), the brightness changes are obtained for different view angles.  \\
Figure~\ref{fig:inclination01} a) shows the  brighness change with the metallicity for four inclination angles $\alpha = [0^{\circ}, 30^{\circ}, 60^{\circ}, 90^{\circ}]$. It is evident that for metallicity values between $\text{M/H}  = -0.1$ and $\text{M/H} = 0.2$ the brightness change increases with decreasing inclination. With increasing $\text{M/H}$ values the effect of inclination  diminishes. The dependence of the inclination is more complex for metallicities greater than  $\text{M/H} = 0.2$.
Note, that our result for different inclinations of the solar case is in agreement with  previous investigations \citep{1998ApJS..118..239R, 2001A&A...376.1080K, 2014A&A...569A..38S}. Therefore, we conclude that the brightness change of  stars with different fundamental parameters can show different dependences on inclination. This is illustrated in~Fig.~\ref{fig:inclination01}~b), where the brightness change with inclination  is shown for three  metallicity values.  For $\text{M/H} = 0.3$, starting from the equator-on view, the brightness change in the Strömgren \textit{b} filter, (Fig.~\ref{fig:inclination01}~b)),  decreases first with decreasing inclination angle. Then,  at lower inclination angles the brightness change increases again towards a pole-on view.  This is explained in more details in the Appendix~\ref{Section:appendix}, where we discuss the center-to-limb variations (CLVs) of faculae. 
\subsection{Effective Temperature}
\label{subsec:effective_temp}
 \begin{figure}
   \centering
   {\includegraphics[width=1.0\linewidth]{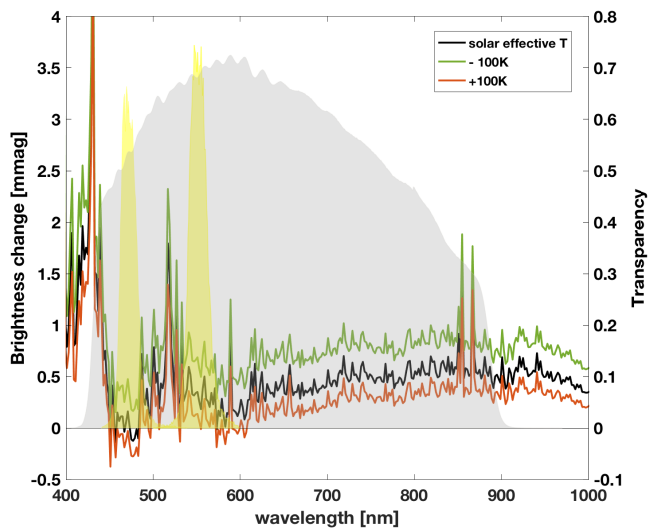}}
      \caption{The amplitude of solar cycle brightness change as defined in Fig.~\ref{fig:model02}. Black line:  The solar case with solar effective temperatures for all features. Green line: A hypothetical Sun, but with $T_{\rm eff}$  reduced by 100K for all components. Orange line: Same as before, but with  $T_{\rm eff}$ increased by 100K for all components. }
         \label{fig:effT01}
   \end{figure}
So far we only studied the effect of  metallicity. Another important  one fundamental stellar parameter that likely affects stellar brightness change is the effective temperature. To understand the role of the effective temperature, the brightness change is calculated for effective stellar temperatures 100 K less  and 100 K greater than the solar value. Such small deviations from the solar value are of special interest, as the measurement accuracy of this parameter is approximately 100 K \citep{2012ApJS..199...30P}. \\
%
 %
The atmospheric models for different effective temperatures were recalculated following the same procedure as described in Section~\ref{Subsec:Radiative_transfer_m}. Subsequently, the brightness change was calculated with SATIRE. Figure~\ref{fig:effT01} shows the total brightness change for the three considered $T_{\rm eff}$ values. Already a 100 K  decrease in $T_{\rm eff}$ causes the brightness change to increase remarkably, while a  similar increase in $T_{\rm eff}$ lowers the brightness change and  leads to spot-dominated cycle in the Str\"omgren \textit{b} filter (for solar $M/H$). Such a sensitive response to a slight change in the effective temperature emphasises the special combination of stellar parameters of our Sun. 
\subsection{Towards Observational Quantities}
\label{subsec:discussion}
 \begin{figure}
   \centering
   {\includegraphics[width=1.0\linewidth]{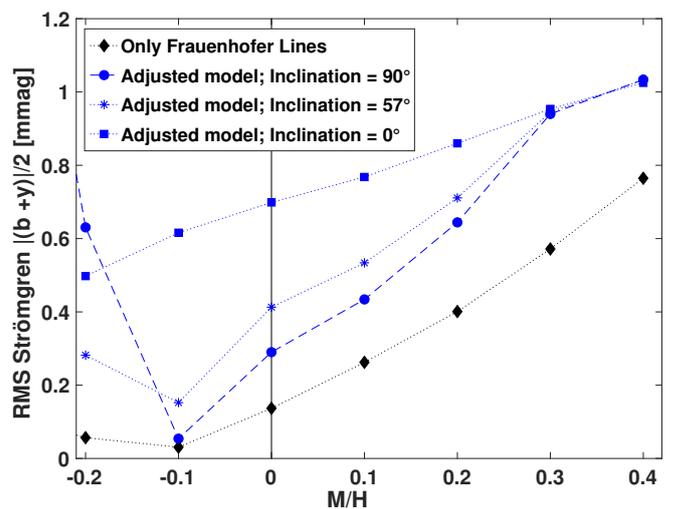}}
      \caption{Estimated rms of the radiative flux change in the Str\"omgren $(b+y)/2$ filters for four different cases vs. metallicity. Black: Only the effect of Fraunhofer line on the opacity is considered. Blue: Recalculated atmosphere models are used.}
         \label{fig:bplusyrms}
   \end{figure} 
Having investigated the effect of metallicity on the spectral brightness change, together with the effect of the inclination, we now  link our comprehensive theoretical results to observed stellar photometric brightness changes. Observed stellar photometric brightness variations is usually analysed in the Str\"omgren \textit{b} and \textit{y} filters separately or as one averaged quantity over Str\"omgren $(b+y)/2$ \citep{2007ApJS..171..260L, 2013ASPC..472..203L, 2018ApJ...855...75R}. Furthermore, observational studies also quantify the brightness change as the root-mean square (rms) variation of the annual mean magnitudes that are obtained from  long-term observations.  Since, however, for our modelling we consider the brightness difference between the magnetic cycle maximum and  minimum of only one cycle, a one-to-one  comparison cannot be made in a straightforward manner. \\
We  approximate the  rms variation  for  Str\"omgren $(b+y)/2$   by using the relation \mbox{$rms = A/ \sqrt{2}$}, where $A$ is the amplitude of a sine function. Figure~\ref{fig:bplusyrms} shows the  rms variations obtained in this way versus metallicity. Note, that the absolute value of the brightness change was considered, i.e. the phase at which maximum brightness is reached was neglected. Overall, Fig.~\ref{fig:bplusyrms} shows that photometric brightness variation for Str\"omgren $(b+y)/2$  is close to a local minimum for  solar metallicity and inclination. The same holds for solar metallicity and inclination angles up to roughly $60^{\circ}$, as well as for the brightness change in the Str\"omgren \textit{b} filter (see Figs.~\ref{fig:Fraunhofer02}, \ref{fig:Adjust03}, and \ref{fig:inclination01}). The smallest rms variations among the computed cases is found for metallicity of $\text{M/H} =-0.1$. Combining this with the results of the effective temperature study reveals that the photometric brightness variations in the Str\"omgren \textit{b}, and Str\"omgren $(b+y)/2$ filter for the solar case is close to a local minimum for the parameter space of the metallicity and effective temperature.
%

\section{Conclusions}
\label{sec:conclusion}
Physics-based models are of importance for a comprehensive understanding of long-term stellar and solar brightness variations, but have so far been rarely applied to stars other than the Sun. Of special interest is the long-standing puzzle that solar brightness variability on the time-scale of the 11-year activity cycle appears anomalously low in comparison to variability of Sun-like stars with a near-solar level of magnetic activity.  Such models are also of importance for the detection of  extra solar planets  \citep[see for example][]{2015A&A...581A.133B}. Thus, there has been a drive towards understanding the solar-stellar connection, especially after the launch of  the Corot  \citep{Borde_Rouan_leger_2003, 2006cosp...36.3749B}  and Kepler \citep{2010Sci...327..977B} space missions that provide broadband stellar photometry of unprecedented precision. In addition, detailed models of stellar variability are of interest for the upcoming TESS \citep{2015JATIS...1a4003R, 2017EPJWC.16001005L} and PLATO \citep{2009ExA....23..329C, 2016AN....337..961R} missions.\\ 
In this study we have extended the well-established solar variability model (SATIRE) to stars with different fundamental parameters. For that we kept the distribution of magnetic features fixed, but we calculated the emergent intensities for different values of metallicity and effective temperatures. In a first step, we demonstrated that changing metallicity affects the Fraunhofer lines in quiet stellar regions in a different way than in faculae. In particular, we find that higher metallicity values result in a significant increase in contrasts of faculae, i.e.~bright magnetic features.  The enhanced contrasts lead to a greater amplitude in the photometric brightness change over a magnetic activity cycle. 
While isolating the effect of Fraunhofer lines on the brightness change confirmed their important contribution, which was first  established for the solar case \citep{2015A&A...581A.116S}, it is crucial to account for the back-reaction of the changed radiation field on the atmospheric structure. In a self-consistent approach, the brightness change is affected by metallicity, such that even small changes  in metallicity values have a significant impact on stellar brightness change for pass-bands used in space- and ground-based stellar observations.  Furthermore, examining  the brightness changes for a hypothetical Sun with slightly changed effective temperature in both directions reveals an increase in brightness change for the Str\"omgren {\textit{b}} filter. \\
All in all we conclude that the combination of the solar fundamental parameters corresponds  to a local minimum in  the brightness change in the Str\"omgren filters on the solar cycle time-scale.  This finding thus explains  the anomalously low solar brightness change by the incidental combination of fundamental solar parameters \citep{2016A&A...589A..46S}. This is a plausible explanation for low solar brightness change. In addition,  a possible observational bias  hinders the identification of stars with low brightness change, and thus potential solar twins. \\
We also find that  the inclination does not have a strong impact on the brightness change for stars with metallicities somewhat higher than solar.   Due to the dependence of  center-to-limb variations on  metallicity, a star with double the metallicity of the Sun would show almost no difference in the brightness changes when it is observed  equator-on  or pole-on. In contrast,  inclination angles  play an important role for stars with low metallicity, i.e.~approximately $\text{M/H} = 0.2$ or lower.  For stars with less metals than in the Sun, different inclination angles can even lead to a transition from faculae-dominated to spot-dominated brightness changes. While this result confirms previous investigations on the importance of the inclination for  solar metallicity, it additionally reveals that the inclination effect becomes weaker for stars with greater metallicities.\\
In the future, our theoretical findings will be tested against observational data.  Recently, sufficient observational data were obtained for one Sun-like star to perform an extensive analysis: The star HD 173701, whose metallicity is twice as large as the solar value, shows higher chromospheric variation, but an even higher photometric brightness variation. Both effects can be explained by the difference in the metallicity \citep{0004-637X-852-1-46}. Another curious case is the Sun-like star HD 143761, which has a near-solar effective temperature but whose metallicity is half of the solar one \citep{2014MNRAS.438.2413V}. This star shows a photometric variability that is spot-dominated, despite being less active than the Sun \citep{2018ApJ...855...75R}. Such a `rule-breaking' behaviour can be explained by a lower facular contrast due to its low metallicity, which is in line with our results. Note, however, that this example should be treated with caution since observational data with two different instruments detected a change from correlated to anticorrelated behaviour \citep{2018ApJ...855...75R}.  Unfortunately, for more detailed comparison we lack a complete set of measurements to determine fundamental parameters for many observed main-sequence stars. Therefore,  an effort is currently underway to obtain a more complete set of observations, including accurate measurements of stellar fundamental parameters and long-term variability for an extended number of Kepler stars \citep{1538-3881-154-3-107, cyKonginprep},  for comparison between modelling and measurements.\\
We conclude that the complex interaction between radiation and matter is crucial to obtain correct brightness variation calculations. However, since the model adjustment uses simplified assumptions and cannot account for three-dimensional effects, 3D MHD calculations are needed for a more realistic approach. Such simulations can provide a more realistic modelling for their brightness changes using a  1.5-dimensional approach \citep{2017A&A...605A..45N}.
Unfortunately, current 3D MHD calculations are only available for different effective temperatures on a coarse grid \citep{2015A&A...581A..42B}.  Consequently, we aim to obtain  3D MHD simulations on a finer effective temperature grid together with different metallicity cases   and to compute the entire spectrum in order to study the dependence of stellar brightness changes on the fundamental stellar parameters.\\
%
\begin{acknowledgements}
This work has received funding from the European Research Council (ERC) under the European Union's Horizon 2020 research and innovation programme (grant agreement No. 715947). This work has been partially supported by the BK21 plus program through the National Research Foundation (NRF) funded by the Ministry of Education of Korea. WS acknowledges travelling support by the grant 200020-~169647 of the  Swiss National Science foundation.
\end{acknowledgements}

%
\bibliographystyle{aa} 
\bibliography{bib} 
%

\begin{appendix} 

\section{Dependence of Brightness Change on Inclination and CLV}
\label{Section:appendix}
\begin{figure}
   \centering
   {\includegraphics[width=0.9\linewidth]{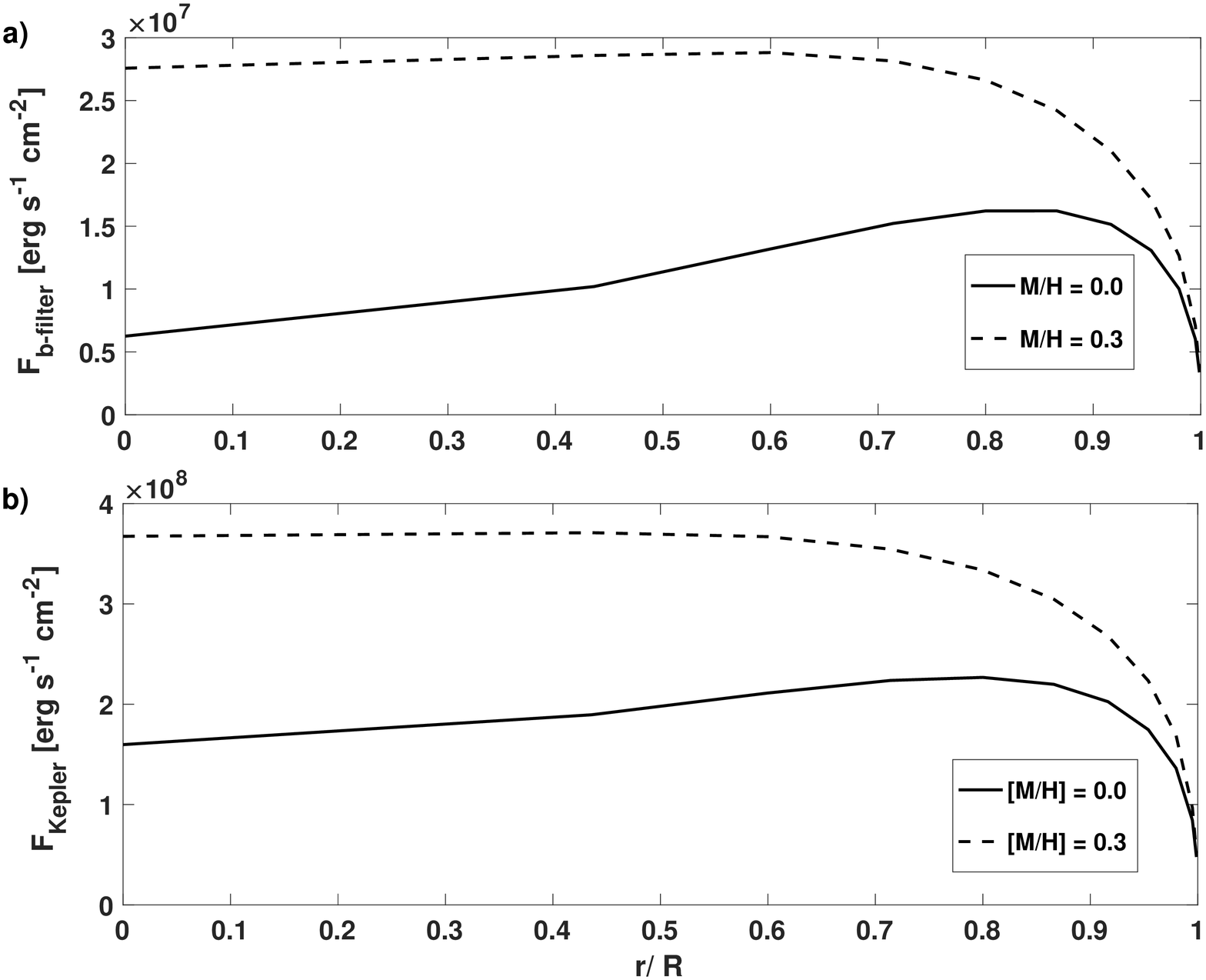}}
      \caption{Center-to-limb variation of facular flux difference.  Here r/R is the normalized radial distance from the centre of the stellar disc. Two flux differences for metallicities are shown; the solar value M/H = 0.0 and  M/H = 0.3. a) Flux differences in the Str\"omgren \textit{b} filter b). In the Kepler pass-band.}
         \label{fig:inclination02}
   \end{figure}
   
\begin{figure}[]
   \centering
   {\includegraphics[width=0.9\linewidth, trim={0.5cm 0 0.0cm 0}, clip]{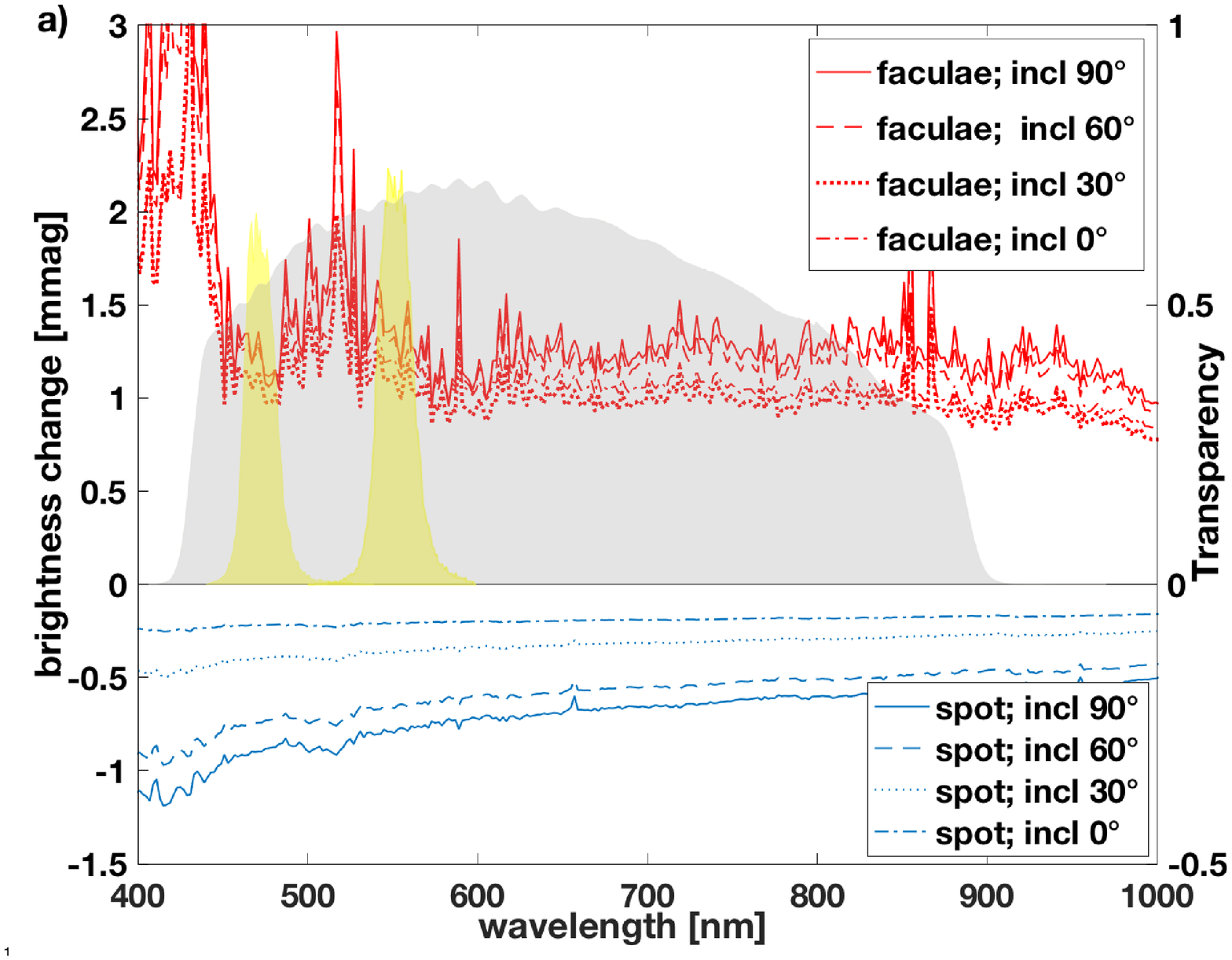}}
     {\includegraphics[width=0.9\linewidth, trim={0.5cm 0 0.0cm 0}, clip]{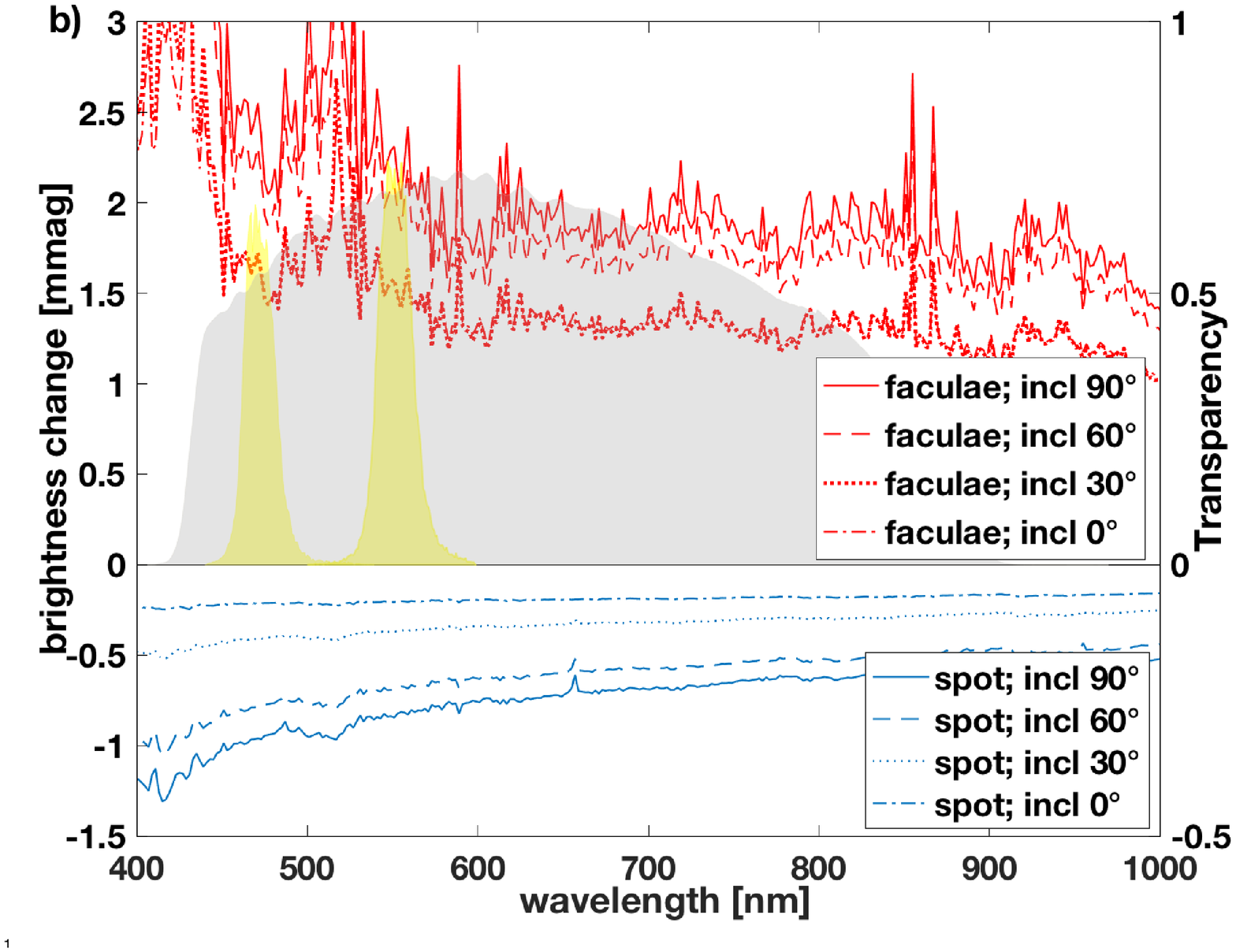}}
      \caption{Brightness change for different components, at different inclination angles for two metallicity values M/H. a) For the solar case $\text{M/H} = 0.0$ the facular (red) and spot (blue)  contributions for different inclinations b) For $\text{M/H} = 0.3$ the facular (red) and spot (blue) contributions for different inclinations. }
         \label{fig:inclination03}
   \end{figure}
To understand why the dependence of brightness change on the inclination itself depends on metallicity, we investigate the center-to-limb variation of facular contrast. 
Figure~\ref{fig:inclination02} shows the CLV of the facular contrast in the Strömgren \textit{b} filter  and the Kepler pass-band for the solar case, and the case with $\text{M/H} = 0.3$. These two cases represent the two different types of behaviour discussed in Subsection~\ref{:subsection:inclination}.  The facular contrast is multiplied by the corresponding $\mu$ to account for the foreshortening effect. 
For the solar case, the facular contrast increases from the disc center outwards.  Since this increase continues almost to the edge of the disc, a relocation of the faclulae towards the center will result in only  a slight decrease of the their contrast with inclination. Thus the effect of the brightness change is weak (see Fig.~\ref{fig:inclination03}~a)).  \\
For  greater metallicity, the facular contrast in the middle of the disc is almost constant, while a steep drop starts at $r/R \approx 0.7$. The difference between the facular contrast around the disc center and the limb is significantly greater than for the solar case. Therefore, two contributions compete in the brightness change when the star with higher metallicity is inclined towards the pole-on view: With inclination some part of the  faculae  are seen closer to the disc center, where the facular contrast does not change, another part is shifted to the limb where the facular contrast drops significantly. Therefore, the facular brightness change decreases greatly with inclination  for a metallicity value of 0.3  (see Fig.~\ref{fig:inclination03}~b)). \\
To explain the different behaviour with inclination for metallicity values in the range $-0.1 < \text{M/H} \leq 0.2$ and metallicity values greater than 0.2, we plot the facule and spot brightness changes with different inclinations for the solar value ($\text{M/H} =0.0$) and the metallicity value $\text{M/H} = 0.3$ in Fig.~\ref{fig:inclination03}. For the solar case, the facular brightness change decreases somewhat with inclination,
at the same time spot brightness change decreases by a much larger amount, so that total brightness change goes up. For the $\text{M/H} = 0.3 $ the brightness change due to spots drops with inclination by a similar amount as for M/H=0, but at the same time facular brightness change drops by a similar amount, or even slightly more, so that the balance between faculae and spots remains either the same or is only slightly shifted towards the spot contribution. Consequently,  the total brightness change either remains the same or decreases. \\
While the total surface area covered by magnetic features remains constant when the star is inclined, due to the equatorial symmetry of the distributions,  the location of magnetic features on the disc changes. Thus the contrasts of the faculae and spots are modified.  Consequently, the total brightness change is altered due  to a changed balance of facular and spot contributions. For the solar metallicity value the faculae contribution decreases slightly with decreasing inclination, i.e.~towards a pole-on view, but at the same time the negative contribution of the spots increases significantly. This leads to an increase of the total brightness change. We find an opposite behaviour for metallicity values greater than $\text{M/H} = 0.2$. For such cases the facular contrast decreases significantly with inclination, while the spot contrast behaves almost as for the solar metallicity value. This results in a decrease of the overall brightness change with inclination. 

\end{appendix}

\end{document}